\documentstyle[newarcrc,fleqn]{article}
\def\sw{SW~Sex}\def\etal{et\,al.}\def\sqig{$\sim\,$}
\def\HeIIl{He\,{\sc ii}\,$\lambda$4686}\def\deg{$^{\circ}$}
\def\HeI{He\,{\sc i}}\def\kmps{km\,s$^{-1}$}

\begin{document}

\title{The SW Sextantis stars}

%\author{Coel Hellier\address{Department of Physics, Keele University, U.K.}}
\author{Coel Hellier\\ [2mm] {\it Department of Physics, Keele University, U.K.}}

\maketitle

\begin{abstract}\noindent 
I review the observational properties of \sw\ stars. I show that they
can be explained by an accretion stream overflowing the disc, combined 
with an accretion disc wind. I suggest that \sw\ behaviour is caused 
by episodes of very high mass transfer, which are balanced by VY~Scl 
low states. 
\end{abstract}

\vspace*{-80mm}

\noindent {\sf Invited review to appear in the proceedings of the Warner Symposium
on Cataclysmic Variables, \\ {\it New Astronomy Reviews}, 1999,
eds P.A.~Charles, A.R.~King \&\ D.~O'Donoghue}
\vspace*{75mm}

\section{Introduction}\noindent 
Thorstensen \etal\ (1991a) coined the term ``SW Sex stars'' to
describe a subset of novalike variables which (1) were typically
eclipsing systems with orbital periods of 3--4 hrs; (2) showed 
distorted emission line wings; and (3) had line-core absorption near
phase 0.5. Other typical characteristics include (4) single-peaked
lines, particularly \HeIIl; (5) peculiar eclipse profiles in the lines
and continuum, and (6) tomograms bright in the lower-left quadrant. 
The same stars also often show VY~Scl low states and superhumps.
See Table~1 for a list of certain and likely class members. 

In this review I'll discuss whether each of these features can be
explained by the suggested models. I report my own
judgement of the state of play, but warn the reader that other authors
might have written a very different review (see Horne 1999).
Finally, I discuss the \sw\ stars as a whole.

\begin{table}[t]
\vspace*{64mm}
\includegraphics{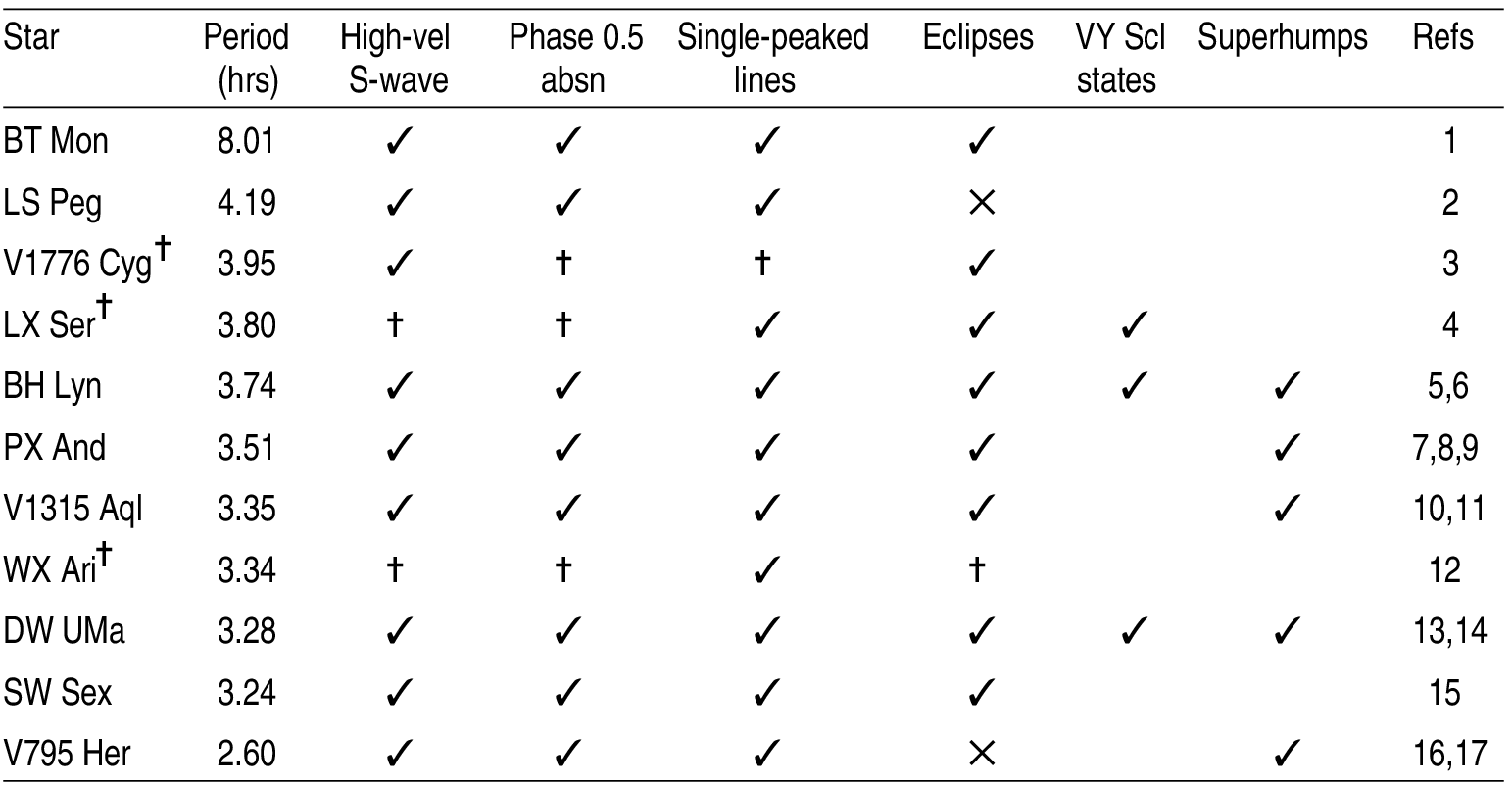}
Table 1: The SW Sex stars. The \dag\ symbol indicates uncertainty.
{\small Refs: Ritter \&\ Kolb (1998); Patterson (1999); 
(1) Smith \etal\ 1998 (2) Taylor \etal\ 1999 (3)
Garnavich \etal\ 1990 (4) Young \etal\ 1981 (5) Thorstensen \etal\ 1991b
(6) Hoard \&\ Szkody 1997 (7) Thorstensen \etal\ 1991a (8) Hellier \&\
Robinson 1994 (9) Still \etal\ 1995 (10)  Dhillon \etal\ 1991 (11) Hellier
1996 (12) Beuermann \etal\ 1992 (13) Shafter \etal\ 1998 (14) Dhillon
\etal\ 1994 (15) Dhillon \etal\ 1997 (16) Casares \etal\ 1996 (17)
Dickinson \etal\ 1997.}
%\caption{The SW Sex stars.}
\end{table}

\section{Emission line wings}\noindent 
Thorstensen \etal's (1991a) paper reminded me of 
EX~Hya in outburst (Hellier \etal\ 1989) where high-velocity
H$\alpha$ wings, crossing from side to side on the orbital cycle, appeared
to result from an accretion stream overflowing the initial impact with
the disc and re-impacting much further in.  Shafter \etal\ (1986) had 
already suggested something similar for \sw\ stars and Lubow (1989)
had given theoretical backing to the idea.
Accordingly, I observed PX~And for 7 orbits with the McDonald 
82$^{\prime\prime}$ telescope --- still one of the best SW~Sex datasets --- 
and computed the velocities of an overflowing stream to produce 
synthetic line profiles. By adding emission from the re-impact point
(with a velocity at the mean of the free-fall stream and the local disc 
velocities) I obtained a good match to the phasing 
and velocity of the observed line wings (Hellier \&\ Robinson 1994; Fig.~1).
Subsequent work has produced similar results in V1315~Aql (Hellier 1996) and 
LS Peg (Taylor, Thorstensen \&\ Patterson 1999). 

In some stars one has to reduce the model velocity to match the data
(e.g.\ to 75\%\ of freefall in V1315~Aql; Hellier 1996).  Justification
for this has subsequently been provided by hydrodynamical modelling of
the stream-disc impact (Armitage \&\ Livio 1998), which shows that after
passing through the impact at the disc rim the infalling stream has a
range of velocities below freefall.

Currently, emission from the stream/disc re-impact appears to be the only
viable model for the distorted line wings. Casares \etal\ (1996) have
criticised it, claiming to see {\it two\/} high-velocity components in
the wings of V795~Her, and have instead invoked magnetic accretion 
to explain them. However, it is clear from their data that the two
features are split by deep absorption, and are more likely to arise from a 
single self-absorbed emission component.
The main difficulty with any model in which magnetic channelling 
dominates the accretion flow is the absence of X-ray emission
in these systems. 

\begin{figure}[t]    % Fig 1
\vspace*{77mm}
\includegraphics{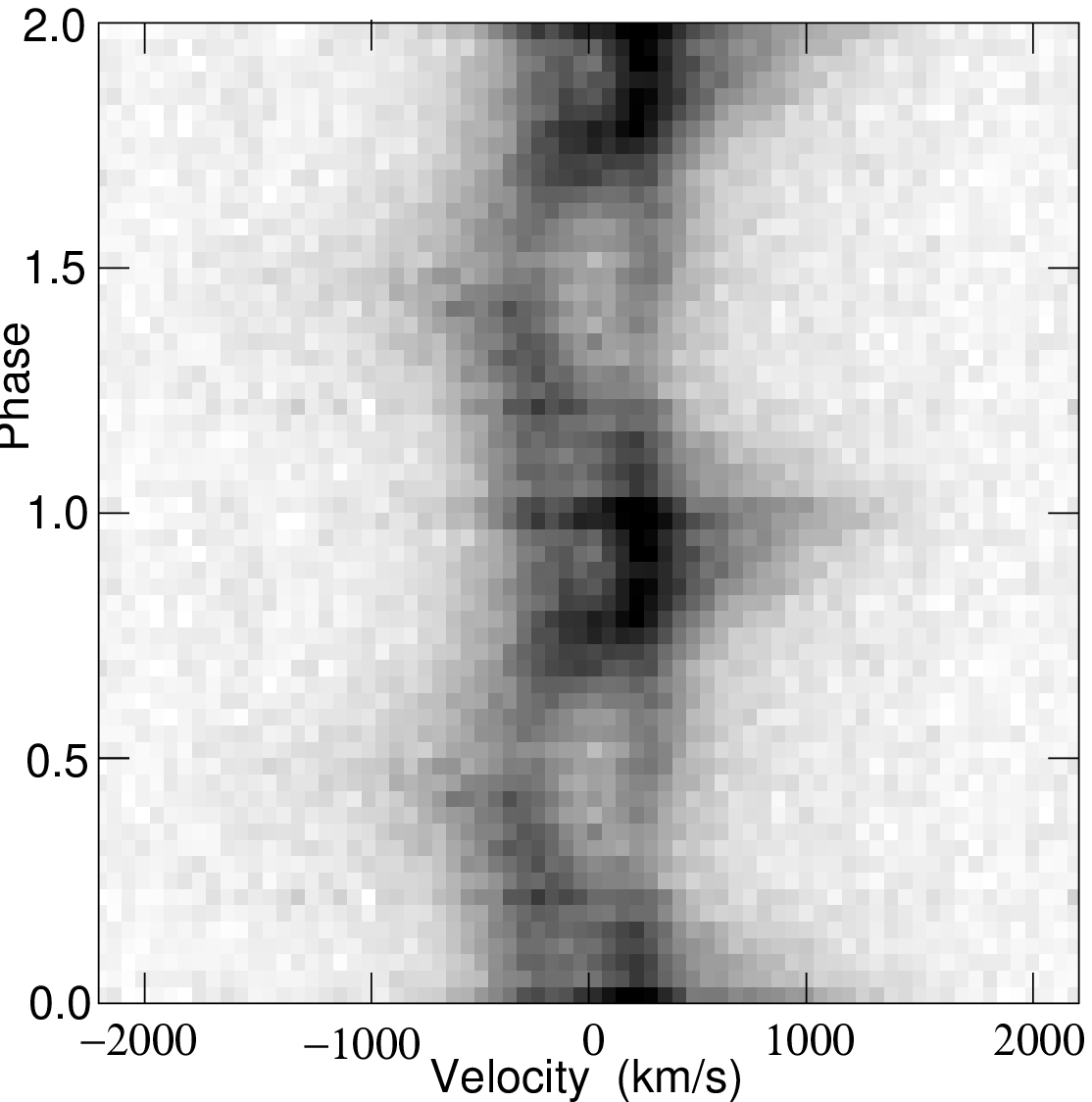}
\includegraphics{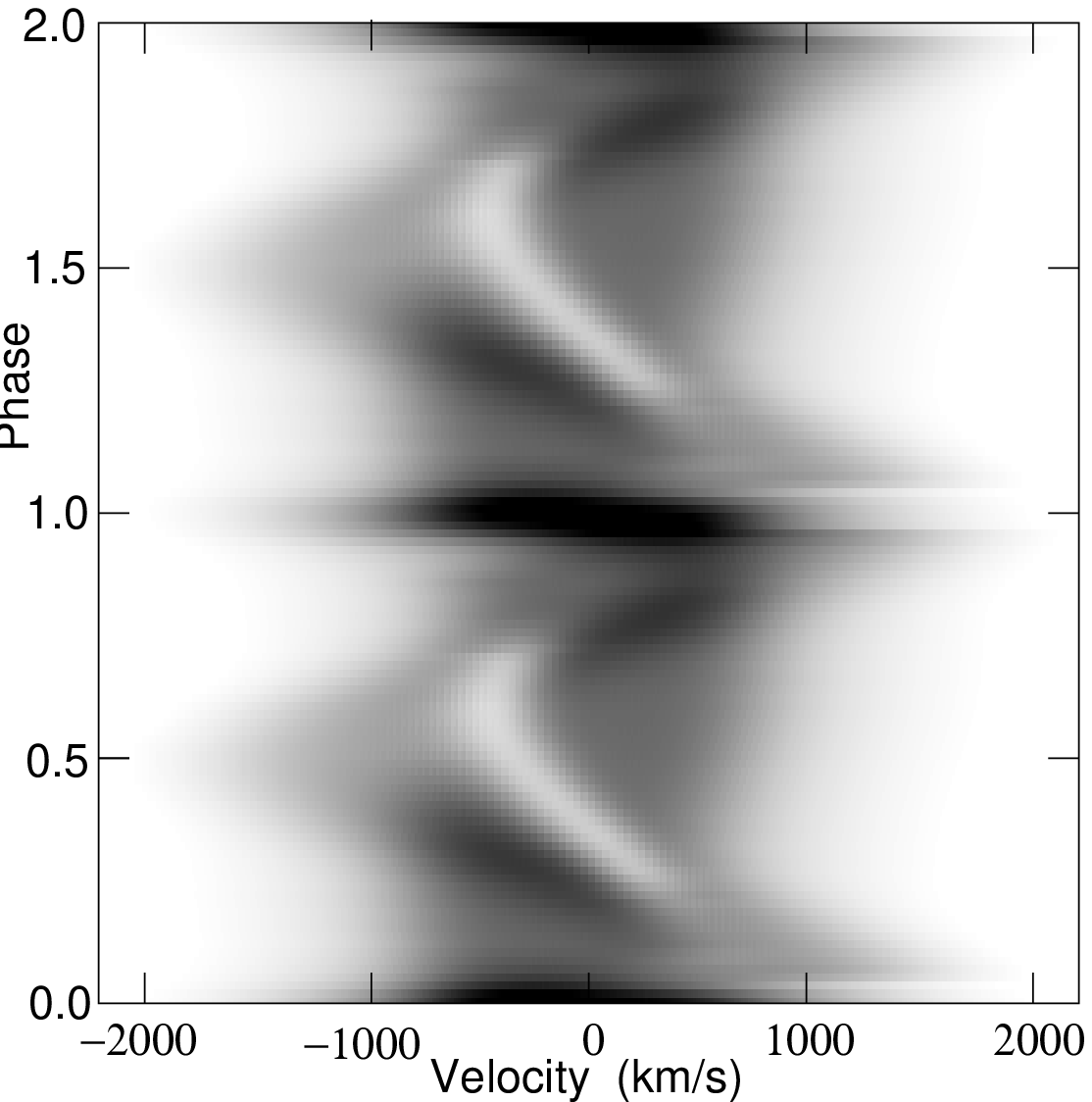}
\vspace*{6mm}\parbox[t]{80mm}{Figure 1: The H$\alpha$ trailed spectra
from PX~And ({\it top left\/}) together with a model simulation 
({\it top right\/}) and the absorption centroid measured by Thorstensen
\etal\ (1991a) from a metal line ({\it right\/}).}\\ [16mm]
\includegraphics{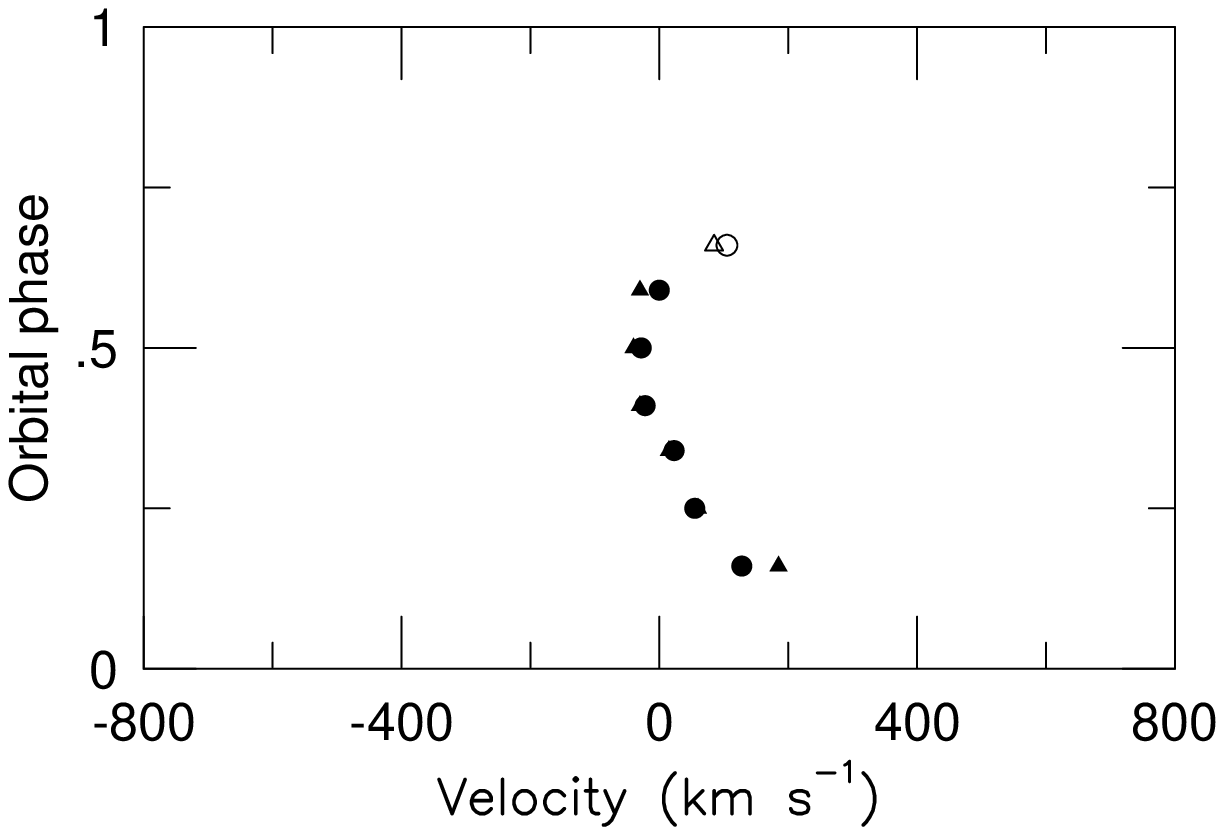}
\end{figure}

\section{The ``phase 0.5'' absorption}\noindent 
When I first computed model spectra I included emission from the full
length of the stream overflowing the disc. This produced an
emission S-wave in the usual place for a stream/disc impact. Looking
at SW~Sex star spectra, however, there is no such feature (e.g.\ Fig.~1);
indeed there appears to be a reduction in flux 
along the track of such an S-wave (max redshift at phase 0.1--0.2;
amplitude \sqig\ 400 \kmps). 

I thus computed models with absorption from the free-fall part of the 
overflowing stream, before the re-impact, ant noticed that it could
additionally explain the phase 0.5
absorption. The first such models (Hellier \&\ Robinson 1994) gave
absorption at all phases, whereas it is observed predominantly 
in the phase range 0.2--0.6. Hence, I invoked a flared disc, so that most
light comes from the rear of the disc in these high-inclination systems. 
The effect is that the stream absorbs the bright rear disc at phases
around 0.5, but has little effect when obscuring the near side of the
disc. A simulation with a 4\deg\ flare in an 82\deg\ inclination system
is shown in Fig.~1 (see also Hellier 1998a). I also show 
the velocity measurements of an absorption-dominated metal line from 
Thorstensen \etal\ (1991a).  There is an excellent match to the phase
range and the velocity trend of the absorption. The velocity amplitude,
though, is too large in the model by a factor \sqig 3. Note, though, 
that (1) I am using free-fall velocities, whereas the material will have
been slowed in the stream/disc bow shock, and (2) the observed absorption
is boxed in by disc and disc-overflow emission, 
biasing the absorption centroid to the line core.  The 
$\gamma$-velocity of the absorption cannot be reliably measured because
of the other components in the Balmer lines, and because the metal line
measured by Thorstensen \etal\ is a blend of uncertain composition. 

In a lower-inclination system the disc flare will have little effect
and the absorption should be seen at all phases. In the non-eclipsing
LS~Peg, Taylor \etal\ (1999) see absorption at
(nearly) all phases in \HeI\,$\lambda$6678 (their fig.~10), and compute a 
tomogram showing absorption in the usual location for a stream/disc impact,
both in accordance with the above model.

The alternatives to explaining the absorption by the overflowing stream
are to invoke disc bulges (e.g.\ Hoard \&\ Szkody 1997) or material
expelled by a propeller (Horne 1999), but neither idea has yet 
been developed enough to make a comparison with the data
similar to that above. 

\section{The single-peaked lines}\noindent 
The most straightforward explanation for single-peaked line profiles,
as seen particularly in \HeIIl, is an accretion disc wind (e.g.\ 
Honeycutt \etal\ 1986; Dhillon \etal\ 1991).
Such profiles have been computed theoretically by Hoare 
(1994). In support of this I have shown that the peak of the \HeIIl\ line
of V1315~Aql moves with the white dwarf motion (Hellier 1996; although
note that the \HeIIl\ wings also contained a disc-overflow component).  
The wind component is also present in the Balmer
lines, filling in any double-peaked disc emission to produce broad,
single-peaked profiles. 

The presence of a wind is confirmed by P~Cygni profiles in the Balmer
lines of V1315~Aql (Hellier 1996; Fig.~2), which are seen at all
phases and appear to move with the orbital motion of the white dwarf. 
Clear wind signatures in optical lines
have also been seen in BZ~Cam (Patterson \etal\ 1996). 

\begin{figure}    % Fig 2
%\vspace*{50mm}
\hspace*{132mm}\parbox[t]{28mm}{Figure 2: P Cygni \linebreak profiles in 
the \linebreak Balmer
lines of \linebreak V1315~Aql.}\\ [15mm]
\includegraphics{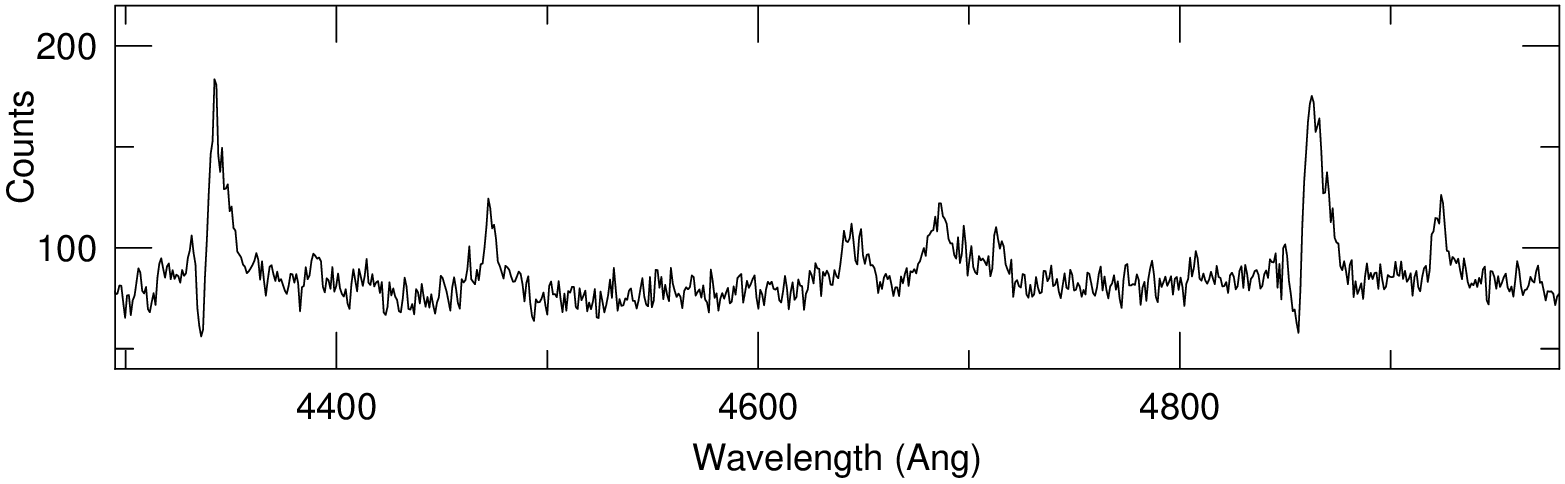}
\end{figure}

\section{The Doppler tomograms}\noindent 
Doppler tomograms of \sw\ lines are often brightest in the lower-left
quadrant, but at a lower velocity than the trajectory of the
stream (a clear example is fig.~13 of Hoard \&\ Szkody 1997;
Fig.~3). This discrepancy has been used to argue against the 
overflow model, and can instead be considered to 
favour the disc-anchored propeller of Horne (1999). If material is 
expelled by a propeller, blobs of different density will collide 
outside the binary, where they would give line emission at the correct 
velocity and phase to explain the tomogram. 

This reasoning, though, assumes that the brightening in the tomogram
can be taken at face value. The emission from the overflowing stream will
have a large velocity dispersion and a mean velocity well below
free-fall (see Armitage \&\ Livio 1998). As pointed out by Hellier \&\ 
Robinson (1994) and Hellier (1996), the brightest region
of the tomogram will occur in the lower-left quadrant where the stream 
component overlaps with the ring of emission from the disc, {\it but this 
brightening is not a component in its own right.\/} The emission in the 
upper-left quadrant is reduced by the absorption discussed above. 

One can see this effect directly in the trailed spectra (e.g.\ Fig.~1). The
high-velocity overflow component zigzags across 
the disc double peaks, causing bright regions where they overlap
(phases 0.35 and 0.85). (The crossings at phases 0.1 and 0.6 are
reduced by absorption along the track of the usual S-wave.)
The relevant region in the tomogram (corresponding to the 
sinusoidal track from +300 \kmps\ at phase 0.85 to --300 \kmps\ at 0.35)
is bright because it links these regions {\it but there is no
discernable component moving along this track.\/} The same effect
is seen in all \sw\ spectra with enough S/N to separate components. 
If the emission were from colliding blobs outside the binary, as in
the propeller model, it would be visible at all phases and so would
produce a continuous S-wave along the track; thus the trailed spectra
do not support the propeller interpretation.

\section{The continuum eclipses}\noindent 
The eclipse profiles of the \sw\ stars do not follow the theoretical
expectation for a novalike disc. The eclipse bottoms are V-shaped, rather
then U-shaped, and when mapped onto a disc produce an inner disc cooler
than expected (Rutten \etal\ 1992). However,
if the above ideas are correct, interpreting the eclipse using a flat
disc will be inappropriate. Instead, one would have to account for (1) a disc
flare; (2) an overflowing, vertically extended stream (which might be brighter or darker than the
novalike disc); and (3) a wind removing energy from the inner disc. 
No study has yet included all these factors. 

A pointer can, however, be gained from recent observations of the
intermediate polar EX~Hya in outburst. These outbursts seem to be 
mass-transfer events in which an enhanced stream overflows the
accretion disc (Hellier \etal\ 1999). They give us a chance to see
an overflowing stream against a faint low-state disc.
The observed eclipse profiles are asymmetric V-shapes, with rapid
ingress and slower egress, and they have minima late by 0.02 in phase
(compared to inferior conjunction of the secondary). These features are
all reproduced by a model eclipse of a bright
stream (Hellier \etal\ 1999; Fig.~4).  The same characteristics
can be seen in the eclipse of \sw\ (fig.~1 of Rutten \etal\ 1992),
although bear in mind that in \sw\ the stream eclipse is convolved with
that of a bright high-state disc.

\begin{figure}    % Fig 3
%\vspace*{50mm}
\hspace*{124mm}\parbox[t]{36mm}{Figure~3: A typical \linebreak \sw\ tomogram 
\linebreak (Hoard \&\ Szkody 1997) \linebreak with a schematic 
interpretation.}\\ [28mm]
\includegraphics{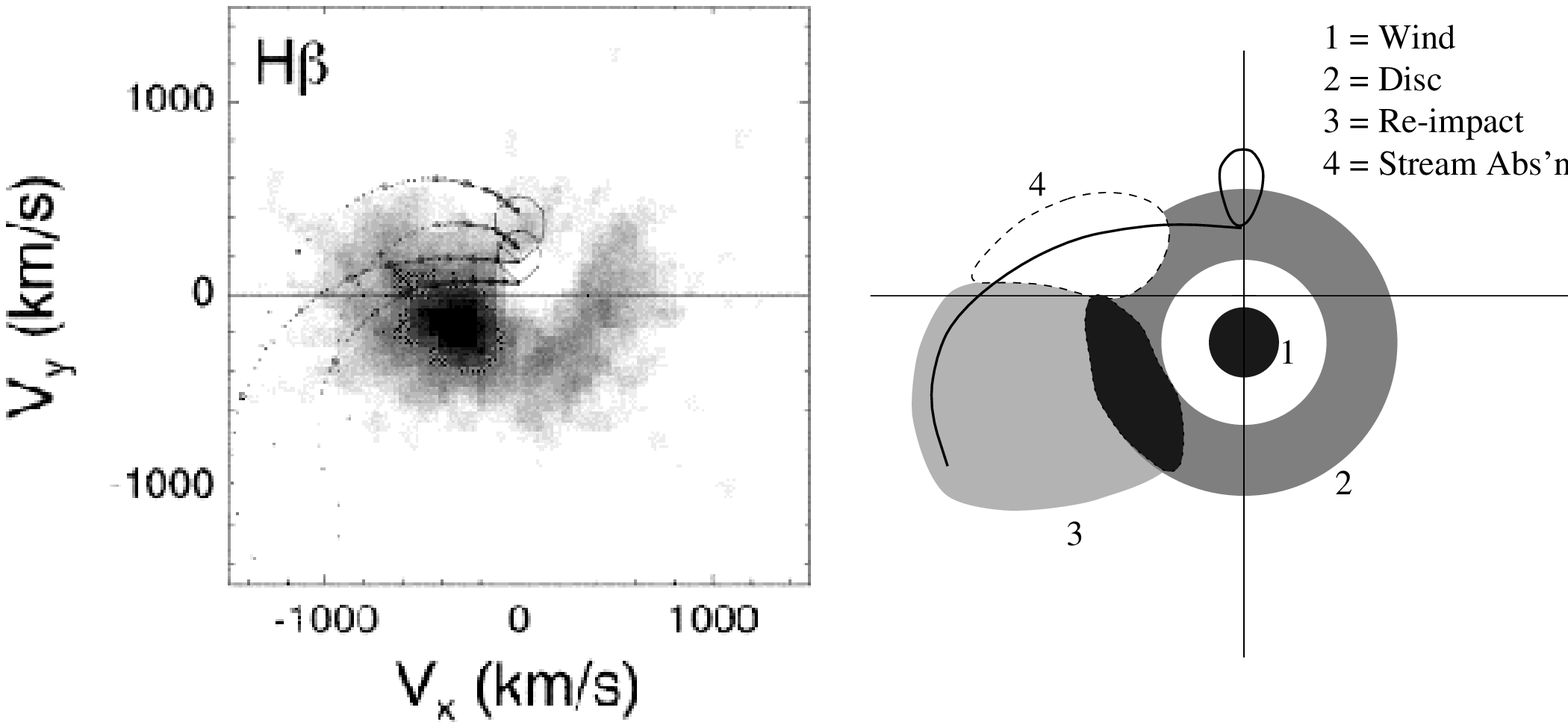}
\end{figure}

\section{The emission-line eclipses}\noindent 
The \HeIIl\ lines are eclipsed to the same extent as the continuum
but the Balmer lines are only partially eclipsed, so that their
equivalent widths increase during eclipse (e.g.\ Dhillon \etal\ 1991). 
This is explained if \HeIIl\ consists of wind emission from 
near the white dwarf where it is deeply eclipsed, and if the Balmer lines
contain a wind component from higher up, where it escapes eclipse. 
The models by Hoare (1994), however, predict the opposite, with Balmer
emission emerging from closer to the white dwarf. Thus either the wind
models are over-simplifed (e.g.\ no clumpiness) or a different idea
is needed. The other evidence for winds (e.g.\ the P Cygni profiles)
suggests the former.

\section{Discussion}\noindent 
I have shown above that the main \sw\ characteristics ---
distorted line wings, distorted eclipses,  
phase 0.5 absorption --- can be explained if the accretion stream 
overflows the disc in these stars. The model, though, is still 
an outline and needs further development, such as the derivation of line 
profiles from the hydrodynamical modelling of Armitage \&\ Livio (1998).

Stream/disc overflow seems to be more widespread than just \sw\ stars. 
For example, the X-ray beat-period pulsations in many intermediate
polars are probably caused by overflow (e.g.\ Hellier 1991, with recent
reviews by Buckley, this volume, and Hellier 1998b). 
Of particular note is FO~Aqr, which shows an X-ray beat
period and also an absorption S-wave from an overflowing stream 
(Hellier \etal\ 1990), reminiscent of \sw\ absorption.

\begin{figure}    % Fig 4
%\vspace*{50mm}
\hspace*{12cm}\parbox[t]{4cm}{Figure~4: Eclipse profiles during EX~Hya's
decline from outburst (crosses and squares) reproduced by a model eclipse
of an overflowing stream (dots).}\\ [15mm]
\includegraphics{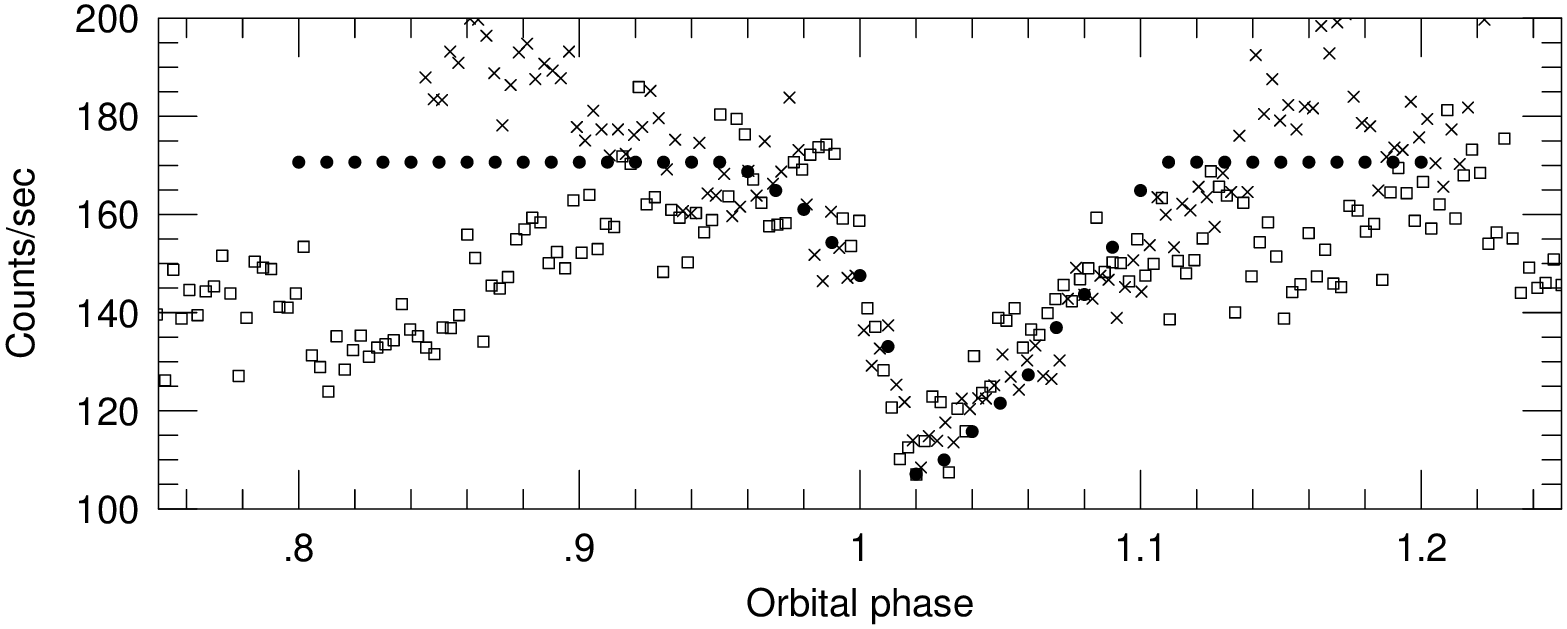}
\end{figure}

Such X-ray beat periods are variable and sometimes absent [pointed
out by Hellier (1991), but now best shown by Wheatley's (1999) dataset
on TX~Col.] The implication of fluctuating overflow probably applies
to \sw\ stars as well. For instance, \sw\ itself has sometimes appeared
much less SW~Sexy than normal (Dhillon \etal\ 1997). Further, we'd expect
there to be novalikes very similar to \sw\ stars, but with little or no
overflow occurring. BP~Lyn and UU~Aqr (Hoard \&\ Szkody 1996; Hoard
\etal\ 1998) seem to be such stars, with (in my judgement) insufficient
phase 0.5 absorption and/or high-velocity S-wave to qualify as fully
fledged \sw\ stars (although there is unlikely to be a clean divide).

So why does overflow appear preferentially amongst a group of
novalikes with 3--4 hr periods?  This is counter to expectations
since theoretically it should be easier to overflow a cool low-state 
disc than a high-state novalike disc (e.g.\ Hessman 1999),
yet overflow has not been reported in quiescent dwarf novae. 

The presence in \sw\ stars of both overflow and winds, and their
tendancy to show superhumps, is circumstantial evidence
for very high mass-transfer rates. Further,  \sw\ stars occur in the
same period range as stars showing VY~Scl low states, and in a range
containing few dwarf novae (e.g.\ Shafter 1992). Thus something appears
to inhibit mass transfer at medium rates. Several mechanism have been
proposed to explain this, including star spots (Livio \&\ Pringle 1994)
and irradiation of the secondary star (Wu \etal\ 1995;
King \etal\ 1996).

Thus if irradiation in a high-$\dot{M}$ novalike leads to even more
mass transfer, it can drive the novalike into an \sw\ state (any similar
mechanism by which mass transfer feeds back into enhanced 
mass transfer would also suffice). When the feedback cycle breaks 
(perhaps through the intervention of star spots or through sheilding
by the accretion disc) the system plunges into a VY~Scl low state.
This mechanism would occur preferentially just above the gap, since the
orbital separation is less than in other novalikes, hence explaining the
period distribution of \sw\ and VY~Scl stars.


\begin{thebibliography}{Armitage, P.\,J., and Livio M.}
\vspace*{-4mm}\noindent\parbox[t]{75mm}{
\bibitem[]{}Armitage P.J., Livio M. 1998, ApJ, 493, 898
\bibitem[]{}Beuermann K. \etal\ 1992, A\&A, 256, 442
\bibitem[]{}Casares J. \etal\ 1996, MNRAS, 278, 219
\bibitem[]{}Dhillon\,V., Marsh\,T., Jones\,D. 1991, MNRAS, 252, 342
\bibitem[]{}Dhillon\,V., Jones\,D., Marsh\,T. 1994, MNRAS, 266, 859
\bibitem[]{}Dhillon\,V., Marsh\,T., Jones\,D. 1997, MNRAS, 291, 694
\bibitem[]{}Dickinson R.J. \etal\ 1997, MNRAS, 286, 447
\bibitem[]{}Garnavich P.M. \etal\ 1990, ApJ, 365, 696
\bibitem[]{}Hoard D.W., Szkody P., 1996, ApJ, 470, 1052
\bibitem[]{}Hoard D.W., Szkody P., 1997, ApJ, 481, 433
\bibitem[]{}Hoard D.W., Szkody P. \etal\ 1998, MNRAS, 294, 689
\bibitem[]{}Hoare M.G. 1994, MNRAS, 267, 153
\bibitem[]{}Honeycutt \etal\ 1986, ApJ, 302, 388
\bibitem[]{}Horne K., 1997, MNRAS, 286, 447
\bibitem[]{}Horne K., 1999, ASP Conf.\ Ser.\ 157, 349
\bibitem[]{}Hellier C., 1991. MNRAS, 251, 693
\bibitem[]{}Hellier C., 1996, ApJ, 471, 949
\bibitem[]{}Hellier C., 1998a, PASP, 110, 420
\bibitem[]{}Hellier C., 1998b, Adv.\ Space Res.\ 22(7), 973
\bibitem[]{}Hellier C. \etal\ 1989, MNRAS, 238, 1107
\bibitem[]{}Hellier C. \etal\ 1990, MNRAS, 242, 250
}\hspace{6mm}\parbox[t]{75mm}{\vspace*{2mm}
\bibitem[]{}Hellier C. \etal\ 1999, in preparation
\bibitem[]{}Hellier C., Robinson, E.\,L., 1994, ApJ, 431, L107 
\bibitem[]{}Hessman F.V., 1999, ApJ, 510, 867
\bibitem[]{}King A.R., Frank J. \etal\ 1996, ApJ, 467, 761
\bibitem[]{}Livio M., Pringle J.\,E., 1994, ApJ, 427, 956
\bibitem[]{}Lubow S.\,H., 1989, ApJ, 340, 1064
\bibitem[]{}Patterson J., \etal\ 1996, AJ, 312, 93
\bibitem[]{}Patterson J., 1999, in `Disc instabilities in close binary
systems', Universal Academy Press, Tokyo, p61
\bibitem[]{}Ritter H., Kolb U., 1998, A\&AS, 129, 83
\bibitem[]{}Rutten R.G.M. \etal\ 1992, A\&A, 260, 213
\bibitem[]{}Shafter A.W., 1992, ApJ, 394, 268
\bibitem[]{}Shafter\,A., Hessman\,F., Zhang\,E. 1988, ApJ, 327, 248
\bibitem[]{}Smith\,D., Dhillon\,V., Marsh\,T. 1998, MNRAS, 296, 465
\bibitem[]{}Still\,M.D., Dhillon\,V., Jones\,D. 1995, MNRAS, 273, 863
\bibitem[]{}Taylor C.J. \etal\ 1999, PASP, 111, 184
\bibitem[]{}Thorstensen J.R. \etal\ 1991a, AJ, 102, 272
\bibitem[]{}Thorstensen J.R. \etal\ 1991b, AJ, 102, 683
\bibitem[]{}Wheatley P., 1999, ASP Conf.\ Ser.\ 157, 47
\bibitem[]{}Wu K. \etal\ 1995, Publ.\ Astron.\ Soc.\ Aust., 12, 60
\bibitem[]{}Young\,P., Schneider\,D., Shectman\,S. 1981, ApJ, 244, 259
}\end{thebibliography}
\end{document}